\documentstyle[aaspp4,psfig]{article}
\tighten
\begin{document}

\title{The Trouble with Hubble Types in the Virgo Cluster}

\author{Rebecca A. Koopmann, Jeffrey D. P. Kenney}
\authoremail{koopmann@astro.yale.edu, kenney@astro.yale.edu} 
\affil{Astronomy Department, Yale University, P.O. Box 208101, New Haven, CT 06520-8101}

\abstract
Quantitative measures of central light concentration and star formation 
activity are derived from R and H$\alpha$ surface photometry
of 84 bright S0-Scd Virgo Cluster and isolated 
spiral galaxies. For isolated spirals, there is 
a good correlation between these two parameters and assigned Hubble types.
In the
Virgo Cluster, the correlation between central light concentration and star 
formation activity is significantly weaker. Virgo Cluster spirals have
systematically reduced global star formation with respect to isolated spirals, 
with severe reduction in the outer disk, but normal or enhanced activity
in the inner disk. Assigned Hubble types
are thus inadequate to describe the range in
morphologies of bright Virgo Cluster spirals. 
In particular, spirals with reduced global star formation activity 
are often assigned misleading early-type classifications, irrespective of 
their central light concentrations. 
45$\pm$25\% of the galaxies classified as Sa in the Virgo Cluster sample have
central light concentrations more characteristic of isolated Sb-Sc galaxies. 
The misleading classification of low concentration 
galaxies with low star formation rates as early-type spirals may account
for part of the excess of `early-type' spiral galaxies in clusters. Thus
the morphology-density relationship is not all due to
a systematic increase in the bulge-to-disk ratio with environmental density.


\section{Introduction}
Discerning environmental effects on the evolution of galaxies depends 
on objective comparisons between cluster and field galaxies.
Traditionally, comparisons are drawn between
galaxies that are assigned to the same Hubble class (e.g., Kenney 1990, 
Oemler 1992). However, the Hubble classification
system was based mainly on nearby field galaxies, and may not
be adequate to describe environmentally altered galaxies in dense 
environments. Attempts to force cluster galaxies into Hubble
type morphological bins may lead to biased conclusions about the physical
differences between cluster and field galaxies, particularly those of
the same assigned Hubble type (see also van den Bergh 1997). Since the Hubble
classification is the framework for the
well-known morphology-density relationship (Oemler 1974; Dressler 1980),
it is especially important to study whether misleading classifications 
contribute significantly to the increase in the fraction of S0's and
early-type spirals (Giovanelli, Haynes, \& Chincarini 1986) in clusters.

As defined by Hubble (1936), classification for spiral galaxies depends on 
three different criteria: 
the relative brightness of the `nuclear' and spiral components (e.g.,
the bulge-to-disk-ratio, the degree of central light concentration), 
the degree to which HII regions are resolved in the spiral arms, 
and the openness of the spiral arms. There is a correlation in the mean 
between the three criteria
for many nearby field spiral galaxies, although the scatter is large
(Kennicutt 1981; Bothun 1982).
Van den Bergh (1976) suggested that application of these criteria in a cluster
environment leads to misleading galaxy classifications,
particularly since the second criterion is strongly dependent 
on the 
star formation rate, which may be systematically affected in clusters 
(e.g., Kennicutt 1983, Moss \& Whittle 1993, Koopmann \& 
Kenney 1998b). Van den Bergh based his DDO 
classifications solely on apparent bulge-to-disk ratios (much as the
Yerkes classification (Morgan 1958) is based primarily on a visual light 
concentration  parameter) and introduced the 
term `anemic' to refer to galaxies with weak star formation.
Bothun (1982) stressed the use of \it quantitative \rm
parameters, such as the bulge-to-disk ratio, to better trace the underlying
stellar mass distribution of the galaxy and avoid dependence on
visually striking star formation characteristics, as well as resolution
effects. Recent studies (Dressler et al. 1994; van den Bergh et al. 1996)
have affirmed that a relatively smaller fraction of distant galaxies
fits the Hubble classification, and 
several groups (e.g., Abraham et al. 1994;
Fukugita et al. 1995; Hashimoto et al. 1998) 
have adopted quantitative parameters, such as central light
concentration, to trace morphological type. 
However, 
no quantitative test of the Hubble classification
has yet been made for nearby cluster galaxies.

In this paper, we explore whether the
Hubble types of nearby isolated and cluster galaxies
correlate with quantitative
measures of the central light concentration and star formation activity.
This work is part of
a series on the comparative star formation properties of Virgo Cluster and
isolated galaxies. A full discussion of star formation properties in the
two samples is given in Koopmann \& Kenney (1998b). 

\section{Observations and Reductions}
The data for the 84 galaxies presented in this paper are extracted from
a larger study of Virgo Cluster and isolated spirals. 
The 55 Virgo S0-Scd galaxies 
have B$_T^0$ $<$ 13 (M$_B$ $<$ -18 for an assumed distance of
16 Mpc), inclinations less than 75$^{\circ}$, and a range in
position in the cluster. Because the main intent was to study the star 
formation properties of Virgo Cluster galaxies, the spiral subtypes are
observed to a relatively high level of completeness, while S0's are incomplete
(Sc: 71\%, Sb: 100\%, Sa: 50\%, S0: 13\%),
and therefore not representative of the Virgo S0 population.
The 29 isolated galaxies were carefully selected from low density regions, 
\it outside groups\rm, as defined by Tully (1987) and
Gourgoulhon, Chamaraux, \& Fouqu\'{e} (1992). All isolated 
galaxies have line-of-sight
velocity $<$ 2000 km s$^{-1}$, M$_B$ $<$ -18, and Tully (1987) density 
parameter $<$ 0.3 gal Mpc$^3$.
Approximately equal numbers of isolated 
S0, Sa, Sb, and Sc galaxies were observed (completeness within selection
criteria for Sc: 23\%, Sb: 25\%, Sa: 67\%, and S0: 63\%);  
thus the relative fractions of the isolated Hubble types
are not representative of the general isolated population, in which S0 and
Sa galaxies are relatively rare. 

Observations were made between 1986 and 1996 at the KPNO 0.9m, CTIO
0.9m, CTIO 1.5m, and WIYN 3.5m telescopes. 
All galaxies were observed in broadband R and in a filter
of bandwidth 70-80 \AA \ appropriate for the redshifted H$\alpha$ line. 
Images were flux 
calibrated using spectrophotometric standard stars, combined
with observations of Landolt standards. 
Continuum-free H$\alpha$ + [N II] (hereafter abbreviated as H$\alpha$), images
were obtained by subtracting scaled R images from the line images. 
R and H$\alpha$ surface photometry and total fluxes were measured using 
procedures in IRAF and with an IDL based surface photometry program (Koopmann
et al. 1998). All surface brightness profiles were adjusted to face on by
assuming complete transparency and a fixed inclination derived from the outer
R isophotes. 
The oversimplified assumption of complete transparency is
necessary in the absence of extinction values for other galaxies, but
should not systematically affect the results since the Virgo Cluster and 
isolated samples span similar inclination ranges.
Images, details of the selection, reduction procedures, and surface photometry
appear in Koopmann et al. (1998) and Koopmann \& Kenney (1998a).

\section{Quantitative Measures of Hubble Criteria}

We measure a central light concentration parameter to trace the
first Hubble criterion. While the bulge-to-disk ratio has been traditionally 
used, measurements of it are dependent
on the models assumed in the decomposition of galaxy surface brightness
profiles, and few galaxy profiles are
well described by the traditional two-component model of an exponential 
disk and an r$^{\frac{1}{4}}$ bulge
(e.g., Freeman 1977; Kormendy 1993; de Jong 1996). 
The degree of central light concentration can be derived in a
model-independent way from surface photometry (e.g., de Vaucouleurs 1977; 
Okamura, Kodaira, \& Watanabe 1984; Kent 1985; Abraham et al. 1994). We define
a concentration parameter similar to Abraham et al. (1994):
$$\rm C30=\frac{F_R(0.3r_{24})}{F_R(r_{24})}$$
where F$_R$(r$_{24}$) is the R flux measured 
within the 24 mag arcsec$^{-2}$ 
isophote, r$_{24}$, and F$_R$(0.3r$_{24}$) is the flux within 0.3r$_{24}$.
Central light concentration parameters have been shown to
correlate with Hubble type (e.g. Abraham et al. 1994), although they may
be less sensitive in differentiating between Sa, S0,
and E galaxies (Smail et al. 1997; van den Bergh 1997). This does not
affect the results of this paper, which
compares differences between S0-Sa galaxies and later-type galaxies. 
The internal measurement uncertainty of C30 
is about 5\% for most galaxies, due to 
uncertainties in the sky background subtraction in R
and inclination uncertainties of $\sim$ 5${^\circ}$. 
Like the bulge-to-disk ratio, the 
concentration parameter is subject to systematic effects which
change the shape of the R profile, including the presence of
newly-formed O and B stars, bright active nuclei, dust, 
or faded outer disks. Half a magnitude of enhancement in
the inner disk or of fading in the outer disk can result in an increase
in C30 of 0.1; a similar amount of extinction in the inner disk can result
in a decrease in C30 of 0.1.

The second criterion in the Hubble classification scheme is the resolution
of spiral arms into HII regions. This criterion is related to the massive
star formation rate (e.g., Larson 1992) in the sense that galaxies which
are actively forming stars will have numerous HII regions along their spiral 
arms. As a tracer of the star formation rate, we measure 
the total H$\alpha$ flux, normalized by the R flux
within r$_{24}$ (hereafter, $\frac{F_{H\alpha}}{F_{R24}}$). Total
H$\alpha$ fluxes were measured using polygonal apertures (within the IRAF
task \it polyphot \rm), with outer boundaries defined by outermost 
HII regions, and are on the
average similar to those of Kennicutt (1983) for 24 galaxies in common.
Values of $\frac{F_{H\alpha}}{F_{R24}}$ correlate well with 
his H$\alpha$ equivalent
width measurements.
The uncertainty in $\frac{F_{H\alpha}}{F_{R24}}$ is typically 20-30\%,
due mostly to uncertainty in the H$\alpha$ continuum subtraction. 

\section{Quantitative Parameters and Hubble Types}
An objective test of
the Hubble classification can be performed 
using the quantitative parameters described above.
For the Virgo Cluster, we adopt Hubble classifications listed by
Binggeli, Sandage, \& Tammann (1985). Classifications
for the isolated galaxies were obtained from 
Sandage \& Tammann (1987) and Sandage \&
Bedke (1994). Although there are discrepancies between the adopted 
classifications and those of de Vaucouleurs et
al. (1991) for individual galaxies, particularly Virgo 
Cluster galaxies, the main conclusions of this paper are not altered 
if these types are substituted (Koopmann \& Kenney 1998b). 

Figure 1 presents the relationship of
the central concentration 
parameter and $\frac{F_{H\alpha}}{F_{R24}}$ in the two
environments. Hubble types
are indicated with different symbols, separated by `whole' class, 
so that Sa stands for Sa-Sab galaxies, Sc for Sc-Scd galaxies, etc.
Lines enclose Sa and Sc galaxies. Several peculiar galaxies, for which
the Hubble classification is obviously inadequate (described further at the
end of this section), are plotted with separate symbols.

For the isolated sample, the
Hubble type is well correlated with both
C30 and $\frac{F_{H\alpha}}{F_{R24}}$.  
Mann-Whitney tests show that differences in the distributions of
C30 and $\frac{F_{H\alpha}}{F_{R24}}$ for different Hubble types
are significant at the 95-99\% level. Furthermore,
there is a good correlation between 
C30 and $\frac{F_{H\alpha}}{F_{R24}}$ (r=-0.74 Spearman $\rho$=-0.79,
significance level 5 x 10$^{-7}$). These results are consistent with
those of 
Hashimoto et al. (1998), who have shown that central concentration
and star formation activity are well correlated for their 
sample of 16377 field galaxies.
The presence of a good correlation between these simple parameters shows
that classification by Hubble type is meaningful for most isolated galaxies.

In the Virgo Cluster, there is a much weaker
correlation between the two quantitative parameters 
(r=-0.45, Spearman $\rho$=-0.47, significance 3 x 10$^{-4}$). The weaker
correlation is primarily caused by a
wider range in star formation activity as a function of concentration,
with both reduced and enhanced levels of activity compared to the isolated
sample. While Virgo Sb and Sc galaxies span similar ranges in C30 as isolated 
Sb and Sc galaxies, the distribution for Virgo galaxies 
classified as Sa is shifted to lower values than isolated galaxies classified
as Sa. The statistical
difference in the distributions in the two environments is at the 98\% level.
This is caused by two effects: 45\% of the Virgo Sa sample have
C30 more typical of isolated Sb-Sc, and the distribution does not extend
to the high concentrations seen in isolated Sa galaxies.
 
Why have galaxies with concentrations more typical of isolated 
later-type galaxies
been classified as Sa in the Virgo Cluster? Virgo galaxies with 
log $\frac{F_{H\alpha}}{F_{R24}}$ flux less than -2.2 tend to be classified 
as Sa, regardless of their central concentrations! In fact, while
Sa and Sb galaxies in Virgo have statistically similar distributions of C30,
they have significantly
different (99\% level) distributions in star formation activity. \it Hence,
the classification of Virgo galaxies as Sa is based more on the
weakness of star formation than on the underlying stellar light distribution.
\rm 
The morphology of these galaxies is therefore not well described by the Hubble
classification, since the stellar mass distribution and star formation
activity are not well correlated. 
The completeness of our Virgo Sa sample is $\sim$ 50\% to B$_T$ $<$ 13; 
thus we estimate that 45$\pm$25\% of bright (B $<$ 13)
Virgo galaxies classified as Sa are actually 
small to intermediate concentration systems with reduced star formation.

The absence of high concentration
Sa galaxies is likely to be at least partially due to the incompleteness
of our sample. Measurements
of concentration for some of the Sa galaxies which are not in our sample are 
listed in other studies
(Okamura et al 1984; Boselli et al. 1997); these numbers 
suggest that there are some higher
concentration galaxies classified as Sa 
in Virgo, but also additional low concentration galaxies classified as Sa.
Given the small number of Virgo S0 galaxies in our sample (13\% complete to
B $<$ 13), direct conclusions about the distribution of C30 for these
galaxies with respect to isolated galaxies cannot be drawn. 

How does the star formation activity in the Virgo Cluster compare to 
isolated galaxies as a function of C30? In the median, 
$\frac{F_{H\alpha}}{F_{R24}}$ of
Virgo Cluster spirals has been reduced by factors of 1.5-3 times 
compared to isolated spirals, 
with the factor increasing towards higher C30. The normalized
H$\alpha$ flux is correlated with HI deficiency: almost all 
the spirals which are HI deficient by at least a factor of 5 fall 
below a log $\frac{F_{H\alpha}}{F_{R24}}$ flux of -2.1. Figure 2 shows
that the reduction in star formation activity has occurred primarily in the 
outer disk, and that the inner
disks of Virgo Cluster spirals have similar or mildly enhanced 
star formation rates. Thus typical Virgo Cluster spirals
with reduced total star formation rates have normal or enhanced
inner star formation
rates, but quiescent outer regions (van den Bergh, Pierce, \& Tully 1990
called this type of morphology `Virgo-type'), rather than disk-wide depression
in the star-formation rate. 

Misleading early-type spiral classifications are also often assigned to
peculiar Virgo Cluster galaxies in an attempt to `shoe-horn' them into
the Hubble classification scheme, analogously to distant galaxies
(van den Bergh 1997). This is the case for several
HI deficient peculiar Virgo spirals with no star formation in their
outer disks, but intermediate central concentrations
and bright H$\alpha$ emission from HII regions within the central
1-2 kpc. Major catalogs assign
these galaxies misleading early-type or `mixed' (Sc/Sa
or Sc/S0) Hubble classifications.  We
designate this type of morphology `St', where St stands for a spiral with
a severely truncated star-forming disk (Koopmann \& Kenney 1998b). 
As shown in Figure 2, the Virgo St galaxies, indicated with filled squares, 
have reduced total and outer disk star formation rates, but normal
inner star formation rates compared to isolated spirals.
If similar galaxies are observed at low resolution
or at large distances, it is likely that they would be classified as
early-type spirals, further contributing to the effect of misleading
classifications on the morphology-density relationship.

\section{Discussion}

We have quantitatively
shown that the weakness of star formation in 
some small to intermediate concentration Virgo Cluster galaxies 
causes them to be assigned early spiral or uncertain
Hubble types, as suggested qualitatively by van den Bergh (1976). 
Cluster galaxies, particularly those with weak
star formation, are not necessarily similar in central concentration
to isolated galaxies with similar Hubble classifications. Data from the
literature suggest that there is evidence for misleading classifications
in other clusters, including Coma (Caldwell et al. 1996; Bothun, Schommer, \& 
Sullivan 1983) and Abell 957 (Abraham et al. 1994). 

To what extent do misleading Hubble classifications contribute
to the morphology-density relationship? While several studies have
concluded that cluster S0 populations cannot be produced by stripped
spirals (Dressler 1980; Boroson, Strom, \& Strom 1983), others suggest they can
(Solanes, Salvador-Sol\'e, \& Sanrom\'a  1989; Eder 1990, Bothun \& Gregg
1990). Our work shows 
that there is a
significant population of low concentration spirals with reduced star formation
in the Virgo Cluster. Thus
the excess of `early-type' spiral galaxies in 
nearby clusters is partially due to misleading classifications of low 
concentration systems with reduced star formation rates, and not all due
to a systematic increase in bulge-to-disk ratio with environmental density.
These results are especially interesting in light of the work of 
Dressler et al. (1997), who 
suggest that as many as 
half the S0 galaxies in nearby clusters have evolved from galaxies which
were spirals at z$\sim$0.5. 
We propose that most of the low concentration `early-type' Virgo spirals 
with low star formation rates
were blue, star-forming, `late-type' spirals a few Gyr ago, and that they 
have recently evolved due to cluster environmental processes. 

What environmental processes have contributed to the evolution of
low concentration galaxies with low star formation rates?
ICM-ISM stripping is likely to play an important role in the removal of gas 
and 
star formation from the outer disk; the Virgo Cluster spiral NGC 4522 is
an example of ongoing stripping (Kenney \& Koopmann 1998). Tidal
interactions between galaxies, which cause the redistribution
of gas inwards and/or outward, also are important in the Virgo Cluster.
At least one of the low concentration Virgo Cluster galaxies with 
reduced outer disk star formation, the St galaxy NGC 4424, is a 
merger remnant, (Kenney et al. 1996) and others show evidence of 
disturbed outer stellar 
disks (Koopmann \& Kenney 1998b). Whether one type of environmental process
is dominant is not yet clear, and at least some galaxies may experience
multiple environmental interactions.

Funding for this work was provided by NSF grant AST-9322779. We are
grateful to Y. Hashimoto, G.D. Bothun, S. Jogee, R.B. Larson, and V.C. Rubin 
for helpful comments.

\newpage

\begin{figure}
\centerline{\psfig{figure=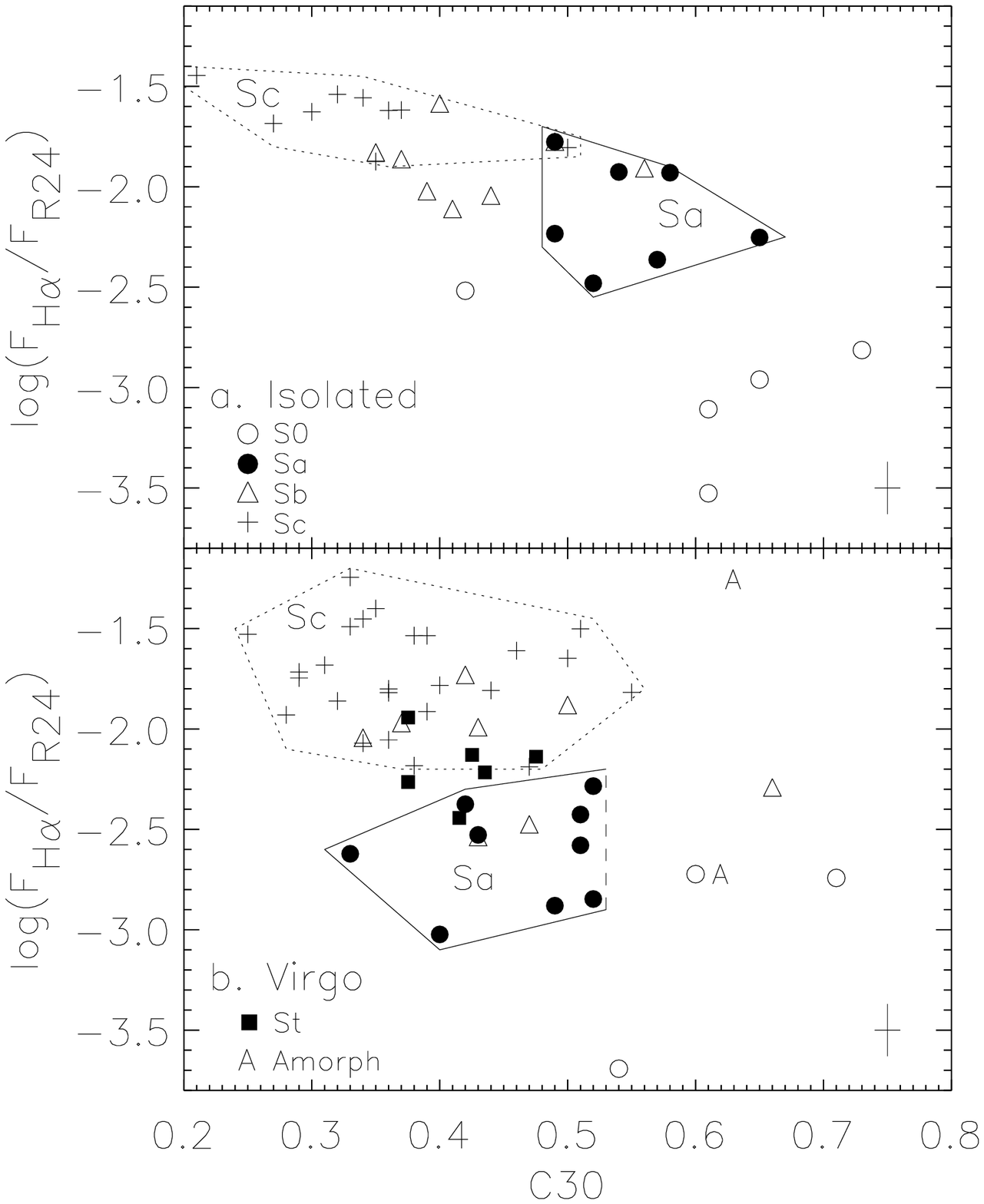,height=6.in}}
\caption{The correlation between C30 and $\frac{F_{H\alpha}}{F_{R24}}$ for
isolated (a) and Virgo (b) galaxies.  Hubble types are indicated by the
symbols. Sa and Sc galaxies are enclosed by
solid and dotted lines, respectively, in each plot. The good 
correlation between the two parameters in (a) indicates that
the Hubble system is meaningful for isolated spirals, while the much
weaker correlation in (b) indicates that the Hubble system is inadequate
for Virgo Cluster spirals. 
In particular, Virgo Cluster galaxies with weak star formation tend
to be classified as Sa galaxies, regardless of their central concentrations.}
\label{fctot}
\end{figure}

\newpage
\begin{figure}
\centerline{\psfig{figure=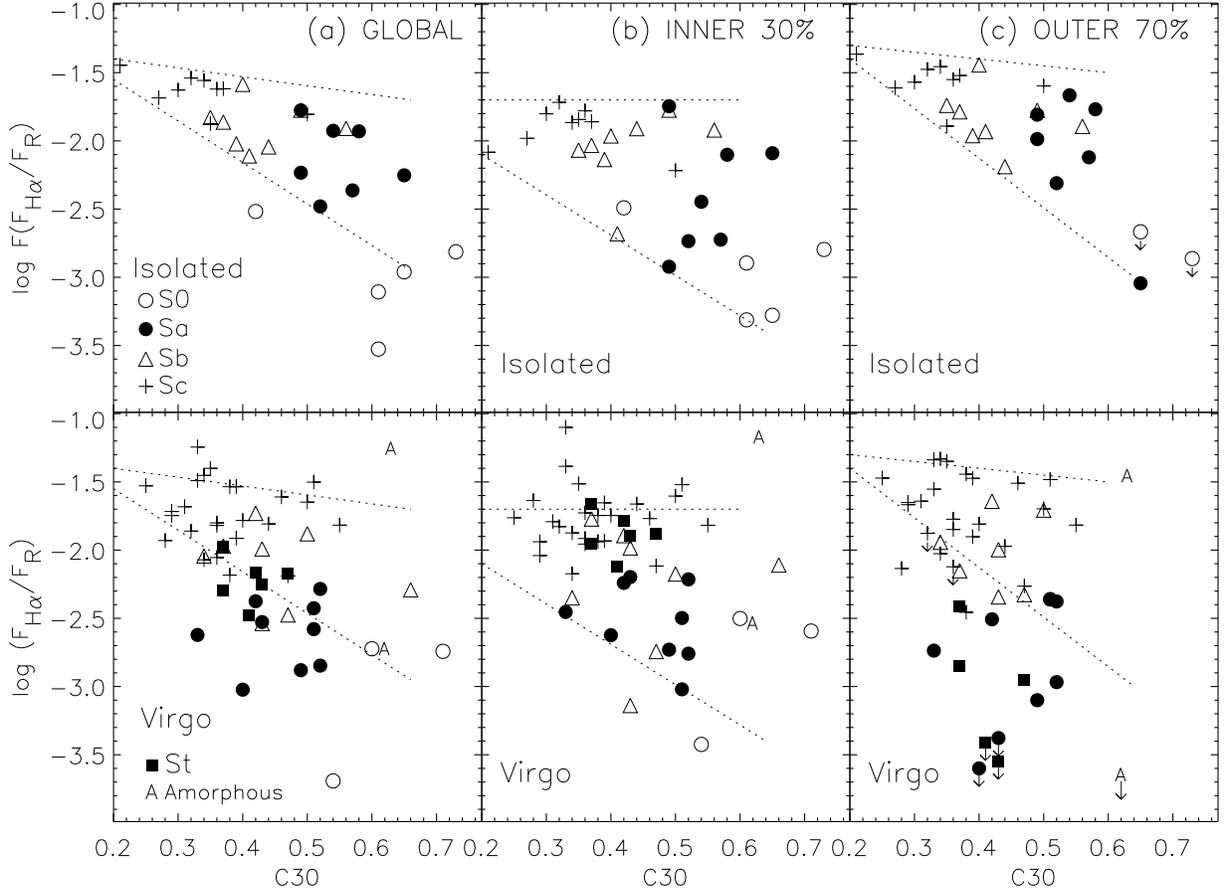,angle=90,height=5.in}}
\caption{H$\alpha$ to R flux ratios as a function of central concentration
for isolated (top) and Virgo (bottom) galaxies: (a) global (same as Figure 1)
(b) inner 30\% of disk and (c) outer 70\% of disk. The dotted lines
indicate the envelope of values for isolated spirals. The inner
star formation activity in Virgo galaxies is similar to or enhanced compared
to isolated galaxies, and the global reduction in star formation activity
in Virgo galaxies is caused by reduced star formation in the outer disk.}
\end{figure}
\end{document}